\documentclass[%
superscriptaddress,
twocolumn,
amsmath,amssymb,
aps,
pra,
floatfix
]{revtex4-2}
\usepackage{amsmath}
\usepackage{amssymb}
\usepackage{graphicx}
\usepackage{graphics}
\usepackage{dcolumn}%
\usepackage{epsfig}
\usepackage{bm}
 
\usepackage{color}

\newcommand{\cK}{{\cal K}}

\newcommand{\bPsi}{\bm{\Psi}}

\newcommand{\bpsi}{ \mbox{\boldmath$\psi$\unboldmath}}

\newcommand{\bvarphi}{ \mbox{\boldmath$\varphi$\unboldmath}}

\newcommand{\bu}{{\bm u}}

\newcommand{\bw}{{\bm w}}

\newcommand{\rev}[1]{\textcolor{black}{#1}}

\begin{document}

\title{Localization of ultracold atoms in incommensurate  spin-orbit-coupling and  Zeeman lattices}

\author{Dmitry A. Zezyulin}
\affiliation{ITMO University, St. Petersburg 197101, Russia}

\author{Vladimir V. Konotop}
\affiliation{Departamento de F\'{i}sica and Centro de F\'{i}sica Te\'orica e Computacional, Faculdade de Ci\^encias, Universidade de Lisboa, Campo Grande, Ed. C8, Lisboa 1749-016, Portugal}

\date{\today}

\begin{abstract}

{\color{black} We consider a particle governed by   a one-dimensional Hamiltonian in which artificial periodic spin-orbit coupling and Zeeman lattice have incommensurate periods.  Using best rational approximations to such quasiperiodic Hamiltonian, the  problem is reduced to  description of spinor states in a superlattice. 
In the absence of a constant Zeeman splitting, the system acquires an additional symmetry, which hinders  the localization. However, if the lattices are deep enough, then localized  states  can appear even for Zeeman field with zero or small mean value. 
Spatial distribution of localized modes is nearly uniform and is directly related to the topological properties of the effective superlattice: center-of-mass coordinates of modes are determined by Zak phases computed from the superlattice band structure.  The best rational approximations feature   the `memory' effect:  each rational approximation holds the information about the energies  and spatial distribution of the modes obtained under preceding, less accurate approximations. \rev{Dispersion of low-energy initial wavepackets is characterized by the law $\propto t^\beta$ with $\beta$   varying between $1/2$ at the initial stage and $1$ at longer, but still finite-time, evolution.} The dynamics of initial wavepackets, exciting mainly localized modes, manifests quantum revivals.
} 

\end{abstract}
\maketitle

\section{Introduction}

Eigenstates of a one-dimensional quantum particle in a potential characterized by two  incommensurate spatial periods, alias in a quasiperiodic potential, are dominated by (but not limited to) spatially localized and delocalized wave-functions that correspond to different regions of 
the energy spectra separated by a threshold energy usually referred to as a mobility edge (ME)~\cite{Mott}. Over the last four decades, properties of quasiperiodic potentials were broadly explored using both the tight-binding approximation, i.e., discrete models (see e.g.~\cite{AA,SoukEkon,Kohmoto,Grempel,Thouless}), and the spatially continuous Schr\"odinger equation with incommensurate potentials (see e.g.~\cite{Azbel,Diener,Boer,Modugno,Biddle,Sarma,Palencia}). The existence of localized and delocalized states can be observed also in two-component systems, like spin-orbit-coupled (SOC) cold atoms~\cite{Galitski13} with spinor components loaded in identical quasiperiodic optical lattices~\cite{ZPZ2013,Adhikari,LocDelocSoc}. It has been established that threshold lattice parameters at which the localization-delocalization transition occurs can be strongly modified by the SOC-induced band flattening~\cite{Yongping,BO,AbdSal}. Emergent phases induced by a uniform SOC in a quasiperiodic tight-binding system on a square lattice have been addressed too~\cite{KohTob,Sahu21}. 

A setting with SOC atoms allows for an essentially novel formulation of the localization problem. First, the SOC itself can be modulated in space~\cite{tunableSpielman,tunableSOC} and, in particular,   be periodic. Second, the components of a spinor describing a SOC atom can experience different, and, in particular, out-of-phase, periodic potentials~\cite{Zeeman-latt} constituting a Zeeman lattice. Even without the SOC modulation, such a lattice affects the dynamics of cold atoms very differently in comparison with conventional optical lattices (see e.g.~\cite{BO}). Study of a simultaneous effect of periodic SOC modulation and a Zeeman lattice, when both have periods whose relation approaches an incommensurate number, is the main goal of the present work. {\color{black} We demonstrate that this previously unexplored system   features a number of  interesting   properties.     First, we find that  
a constant component of the Zeeman splitting plays a prominent  and ambivalent role: when it is absent, the system acquires an additional symmetry, which imposes that any localized state must be degenerate and  two-peaked.  As a result, in this case the localization requires deeper lattices, than in the case where the Zeeman field has nonzero mean. 
At the same time, at the constant Zeeman splitting large enough all modes become delocalized.  As a result, there exists a parametric region, where the most pronounced localization is achieved for intermediate values of the constant Zeeman field.  Second, using the best rational approximations to the incommensurate lattices, we reduce the problem to a spinor in a periodic superlattice and   uncover the relation between spatial distribution of localized modes and topological properties of the effective superlattice. Namely, the coordinate of the center-of-mass (COM) of each localized mode is determined by the Zak phase~\cite{Zak} of the respective superlattice miniband. Third, we report the `memory' effect for  successive best rational approximations: each approximation has the memory about the preceding, i.e., less accurate, ones. Finally, studying the dynamics of initially localized wavepackets we observe the oscillatory behaviour interpreted as a signature of quantum revivals.}

The paper is organized as follows. In Sec.~\ref{sec:model} we introduce the model and describe our approach which relies on the approximation of the quasiperiodic two-component Hamiltonian with an exactly periodic superlattice. Section~\ref{sec:results} presents numerical results on the localization of the eigenstates of the obtained Hamiltonian. Section~\ref{sec:dynamics} addresses the dynamics of the system below and above the localization transition. Section~\ref{sec:concl} concludes the paper.

\section{The model}
\label{sec:model}

\subsection{Best rational approximations and effective Hamiltonians}

Let us consider a SOC atom governed by the dimensionless Hamiltonian as follows 
\begin{align}
	\label{hamilt}
	 H=\frac{1}{2}\left[-i \partial_x  + a(x)\sigma_1 \right]^2+\frac{\Lambda+ \Omega(x)}{2}\sigma_3.
\end{align}
Here $a(x)$ is a real-valued $\pi$-periodic function (it will  be referred below as a SOC lattice): $a(x)=a(x+\pi)$ and $\sigma_{1,2,3}$  are the Pauli matrices. The Zeeman field consists of a constant component $\Lambda$ and a $\pi/\kappa$-periodic  lattice: $\Omega(x)=\Omega(x+\pi/\kappa)$ where $\kappa$ is an irrational number, i.e., the periods of the SOC lattice and Zeeman lattice are incommensurate.

Since a real-world atomic system is finite and, without loss of generality, can be centered at $x=0$,   
we address the eigenvalue problem on a finite interval considered sufficiently large
\begin{align}
	H\bpsi=E\bpsi, \qquad x\in[-\ell/2,\ell/2],
\end{align}
where  $\ell$ denotes the   spatial extent  of the system.

For any irrational  $\kappa$ there exists a sequence   $\{\kappa^{(N_1)}, \kappa^{(N_2)}, \ldots\}$  of the best rational approximations (BRAs) of gradually improving accuracy  (see e.g. \cite{Khinchin}). Here `the best' means that if a fraction $\kappa^{(N)} = M/N$ is one of the BRAs (with $M$ and $N$ being coprime integers), then it approximates $\kappa$ better than   any other  rational number    with the denominator less or equal than $N$. More formally, $\kappa^{(N)} = M/N$ is one the BRAs, if for any pair of coprime integers $P$, $Q$, such that $M/N\ne P/Q$ and $0<Q\leq N$, one has 
$\left| Q\kappa - P  \right | > \left| N\kappa -  M \right | $. The sequence of  BRAs can be constructed from   the continued  fraction associated with the  irrational number $\kappa$: truncation of the infinite continued fraction to a finite number of terms yields one of the BRAs. The more terms one keeps in the truncated continued fraction, the better accuracy of the obtained BRA.
 
Let $\kappa^{(N)}=M/N$ be one of the BRAs to $\kappa$ (hereafter we use upper index $N$ in order to refer to the BRA with denominator $N$; then the numerator $M$ is uniquely defined). Define a $\pi/\kappa^{(N)}$-periodic function $\Omega^{(N)}(x)=\Omega^{(N)}(x+(N/M)\pi)$, obtained from $\Omega(x)$ by the replacement $\kappa\to\kappa^{(N)}$ and introduce the respective Hamiltonian $H^{(N)}$
\begin{align}
	\label{hamilt-N}
	H^{(N)}=\frac{1}{2}\left[-i \partial_x  + a(x)\sigma_1 \right]^2+\frac{\Lambda+ \Omega^{(N)}(x)}{2}\sigma_3.
\end{align}
Hamiltonian (\ref{hamilt-N}) features a combination of two commensurate lattices $a(x)$ and $\Omega^{(N)}(x)$ and therefore 
represents a periodic superlattice with a period equal to $L^{(N)}=\pi N$. Assuming that $\Omega(x)$ is a continuously differentiable function, one can always find  a sufficiently accurate BRA in the sense that the difference  
\begin{align}
	 \omega^{(N)}=
	\Omega(x)-\Omega^{(N)}(x)=\mathcal{O}\left((\kappa-\kappa^{(N)})x\right), \quad |x|\leq \frac{\ell}{2}
\end{align}
is as small as necessary for $x\in [-\ell/2,\ell/2]$. Thus, any eigenstate of $H$ localized on the interval $\ell$ can be considered as a weakly perturbed state in the Hamiltonian $H^{(N)}$ on the same interval; notice that above requirement on  $\omega^{(N)}$ implies $L^{(N)}\gg\ell$.

Several differences between the introduced system (\ref{hamilt}) and its approximations (\ref{hamilt-N}), and the previously studied discrete models of the Aubry-Andr\'e (AA) type are to be emphasized. Being of spinor character, the Hamiltonian (\ref{hamilt}) is four-parametric, unlike the two-parametric AA model. With the period of SOC lattice being fixed, these four parameters are the amplitude of the SOC lattice $\alpha$ [see (\ref{eq:a(x)}) below], the amplitude of the Zeeman lattice $\Omega_0$,   the constant Zeeman splitting $\Lambda$, and the phase shift $\theta$ between the two lattices. Furthermore, although the incommensurate limit corresponds to the Brillouin zones of the successive approximations shrinking to zero (see illustration in Fig.~\ref{fig:periodic} below) and thus to the limit of extremely flat low bands, the tight-binding approximation is not applicable. Indeed, in the incommensurate limit the lattices $a(x)$ and $\Omega(x)$ are not required to be deep enough to justify the tight-binding approximation for several lowest levels. Furthermore, a Zeeman lattice incorporates sublattices for two spinor components having opposite signs: where for one component a potential has a  minimum, the potential for the other component has  a maximum and \textit{vice versa}. Thus, the localization of the modes below the ME is not determined by positions of equally spaced deep potential minima; instead, as we show below, the places where modes are situated are determined by the Zak phases, i.e., by the topology of the superlattice.   

\subsection{Periodic boundary conditions}
 
To describe the localized states of $H$, one can consider the eigenvalue problem for $H^{(N)}$ (c.f.~\cite{Diener,Modugno}) subject to desirable boundary conditions. Periodicity of   $H^{(N)}$ implies that periodic boundary conditions are the most promising choice, whose advantages include a possibility to relate the problem to  the Bloch theory and employ topological characteristics of periodic systems. 
Therefore from now on our goal is the study of localization of atomic states and evolution of wavepackets governed by the approximated Hamiltonian $H^{(N)}$ with the focus on sufficiently large $N$ (formally tending to infinity). More specifically, we consider the eigenvalue problem for $H^{(N)}$ 
\begin{align}
		\label{eigen}
	H^{(N)}\bpsi^{(N)}(x) =E^{(N)}\bpsi^{(N)}(x), 
\end{align}
where $\bpsi^{(N)}(x)$ is a two-component spinor wavefunction, on the interval
\begin{align}
		x\in[- {L^{(N)}}/{2}, {L^{(N)}}/{2}]=:I^{(N)}
\end{align}
subject to periodic boundary conditions: 
\begin{align}
	\label{pbc}
	\bpsi^{(N)}(- {L^{(N)}}/{2})=\bpsi^{(N)}( {L^{(N)}}/{2}), 
\end{align}
and normalization 
\begin{align}
\int_{I^{(N)}} [\bpsi^{(N)}(x)]^\dagger  \bpsi^{(N)}(x)\,dx =1.
\end{align}

The eigenvalue problem (\ref{eigen}) with boundary conditions (\ref{pbc}) has a discrete spectrum whose eigenenergies will be denoted by  $E_1^{(N)}\leq E_2^{(N)} \leq \ldots \leq E_\nu^{(N)} \leq\ldots$, where the lower index $\nu=1,2,\ldots$   enumerates   the eigenenergies and respective eigenvectors $\bpsi_\nu^{(N)}$.  Given an eigenstate $\bpsi_\nu^{(N)}$, the quantitative   measure of its localization  within the interval $I^{(N)}$  can be conveniently represented
by   the  inverse participation ratio (IPR) 
\begin{align}
\label{IPR}
\chi_\nu^{(N)}= {\int_{I^{(N)}} 
\left([\bpsi_\nu^{(N)}]^\dagger \bpsi_\nu^{(N)}\right)^2 dx}.
\end{align}
Large, $\chi_\nu^{(N)} \gg 1/L^{(N)}$, and small, $\chi_\nu^{(N)}\sim 1/L^{(N)}$, values of the IPR correspond to localized and delocalized states. 

Under the periodic boundary conditions (\ref{pbc}), a position of a localized mode within the superlattice period can be computed as~\cite{Resta}
\begin{equation}
\label{eq:Xp}
\mathfrak{X}_\nu^{(N)} = \frac{L^{(N)}}{2\pi}\arg \left\{ \int_{I^{(N)}} 
[\bpsi_{\nu}^{(N)}]^\dagger  \bpsi_{\nu }^{(N)}   e^{{2\pi ix}/{L^{(N)}}}dx\right\},
\end{equation}
where the principal  value  of the argument must be chosen, i.e.,   $\arg \in (-\pi, \pi]$. Generally speaking, the position defined in (\ref{eq:Xp})  is  different from the conventional COM which is defined as 
\begin{align}
\label{eq:COM}
x_\nu^{(N)}=\int_{I^{(N)}} 
x  [\bpsi_{\nu}^{(N)}]^\dag  \bpsi_{\nu }^{(N)}  dx
\end{align}
and is usually used when localization on the whole real axis is considered. In the meantime, the difference between $\mathfrak{X}_\nu^{(N)}$ and $x_\nu^{(N)}$ becomes appreciable only for    states localized near the boundaries of the   interval $I^{(N)}$, i.e., near $x=\pm L^{(N)}/2$. Since we are interested in the limit $N\gg 1$, the relative number of such modes is small, which means that their contribution to the results presented below is negligible. Therefore  we will have $\mathfrak{X}_\nu^{(N)} \approx x_\nu^{(N)}$ for almost all localized modes. 


\subsection{Superlattice band structure and Zak phases}

Since the Hamiltonian $H^{(N)}$ admits the periodic continuation from the interval $I^{(N)}$ to  the entire real axis, it is natural to explore the band-gap spectrum of the corresponding eigenvalue problem. Considering $x\in\mathbb{R}$
one can formulate the eigenvalue problem as follows: $H^{(N)}\bvarphi_{\nu k}^{(N)}(x)=\varepsilon_\nu^{(N)}(k)\bvarphi_{\nu k}^{(N)}(x)$. Here,  $\bvarphi_{\nu k}^{(N)}(x)=e^{ikx}\bu_{\nu k}^{(N)}(x)$ are Bloch states  with $\bu_{\nu k}^{(N)}(x)=\bu_{\nu k}^{(N)}(x+L^{(N)})$ being $L^{(N)}$-periodic functions, $\varepsilon_\nu^{(N)}(k)$ are the energies, index $\nu=0,1,\ldots$ enumerates the spectral bands, and the Bloch wavenumber $k$ runs over the reduced Brillouin zone of the superlattice: $k\in [-1/N, 1/N)$.  The usual normalization requires that $\int_{I^{(N)}} 
[\bu_{\nu k}^{(N)}]^\dag  \bu_{\nu k}^{(N)} dx = 1$.

If the superlattice spectrum $\varepsilon_\nu^{(N)}(k)$ is computed, the eigenvalues and eigenvectors of the eigenvalue problem (\ref{eigen}) with periodic boundary condition (\ref{pbc}) can be   obtained as  $E_\nu^{(N)}=\varepsilon_\nu^{(N)}(0)$ and  $\bpsi_\nu^{(N)}(x)=\bvarphi_{\nu 0}^{(N)}(x)$.  \rev{In other words, 
each solution of  (\ref{eigen})--(\ref{pbc}) is a  Bloch state at  $k=0$ of the effective periodic lattice defined on the whole real axis by the given BRA}.
Moreover, using the superlattice band structure, a spatial position  of localized modes $\bpsi_\nu^{(N)}(x)$  defined in (\ref{eq:Xp}) can be estimated from   the topology of the corresponding  band. Indeed, for each  
band of the superlattice one can compute the Zak phase  
\cite{Zak} 
\begin{align}
	\gamma_\nu^{(N)} =\int_{-1/N}^{1/N}
\Xi_\nu^{(N)}(k)dk,
\end{align} 
where
\begin{align}
	\label{eq:X}
	\Xi_\nu^{(N)}(k)=i\frac{2}{N}\int_{I^{(N)}} 
	[\bu_{\nu k}^{(N)}(x)]^\dagger \frac{\partial}{\partial k}\bu_{\nu k}^{(N)}(x)dx
\end{align}
is the Berry connection.
In terms of  the Wannier functions~\cite{Kohn}  
\begin{align}
	\bw_\nu^{(N)}(x-\pi m N)&=\sqrt{
	\frac{N}{2}
	}\int_{-1/N}^{1/N}
	e^{-i\pi k mN}\bvarphi_{\nu k}^{(N)}(x) dk,
\end{align}
the Zak phase can be is expressed as
 \begin{align}
	\gamma_\nu^{(N)}
	=\frac{ N}{2}\int_{-\infty}^{\infty}x [\bw_\nu^{(N)}(x)]^\dagger\bw_\nu^{(N)}(x)dx.
\end{align}
Those eigenstates   which are  well localized inside the interval $I^{(N)}$   can be approximated by the  Wannier functions computed from the respective superlattice bands, i.e., $\bpsi_\nu^{(N)}(x)\approx\bw_\nu^{(N)}(x)$ for $x\in I_N$, and  therefore we obtain the relations
\begin{align}
	\label{x-Zak}
	\mathfrak{X}_\nu^{(N)} \approx x_\nu^{(N)} \approx \frac{2}{N}  \gamma_\nu^{(N)}. 
\end{align} 	 

{To conclude this section, we emphasize that  while periodic boundary conditions are used, and the effective periodic extension of the potential is exploited, the physical applications of the results remains  meaningful only inside the interval $I^{(N)}$.}

\section{Numerical results}
\label{sec:results}

\subsection{Rational approximations}

To perform a numerical study, we have chosen the $\pi$-periodic SOC lattice in the form
\begin{align}
\label{eq:a(x)}
a(x) = \alpha \cos(2x),
\end{align}
where $\alpha$ is the lattice depth, and  the $\pi N$-periodic Zeeman lattice
\begin{align}
	\Omega^{(N)}(x)  = \Omega_0 \sin\left(2 \frac{M}{N} x + \theta\right), \label{eq:Omega(x)}
\end{align}
where the fraction $M/N$ is one of the BRAs to $\kappa=\sqrt{2}$.  Respectively, the function $	\Omega^{(N)}(x)$ is an approximation to  $\Omega(x)  = \Omega_0 \sin\left(2 \sqrt{2}  x + \theta\right)$ in the quasiperiodic Hamiltonian (\ref{hamilt}). In the explicit form, several first BRAs for $\sqrt{2}$ are  given  as   \cite{Koshy}  
\begin{equation}
	\label{eq:RA}
	\frac{M}{N}=
	{\frac{1}{1},}\,\, 
	\frac{3}{2},\,\, \frac{7}{5},\,\, \frac{17}{12}, \,\,\frac{41}{29},\,\, \frac{99}{70},\,\, \frac{239}{169},\,\, \frac{577}{408},\,\, \frac{1393}{985},\ldots
\end{equation}
We emphasize however that our analysis, as well as the main conclusions remain valid for any irrational number. Nowhere in the subsequent analysis the specificity of the above choice is used; only the set of the fractions $M/N$ (\ref{eq:RA}) giving the best rational approximation will be modified for another choice of $\kappa$.  A specific choice of the phase shift between the SOC and Zeeman lattices,  i.e., angle $\theta$ in (\ref{eq:Omega(x)}), also has no significant impact on the results presented below, except for certain particular values of $\theta$ at which the system acquires an additional symmetry (see below in Sec.~\ref{sec:double}).

\begin{figure}
	\begin{center}
		\includegraphics[width=0.999\columnwidth]{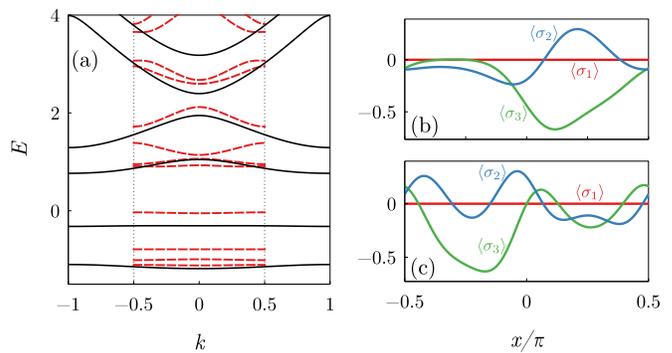}%
	\end{center}
	\caption{ (a) The lowest energy bands for the $\pi$-periodic system with $M=N=1$ (\rev{solid} black lines) and $2\pi$-periodic system with   $M/N=3/2$ (\rev{dashed} red lines) are shown in the respective first Brillouin zones $[-1,1]$ and $[-1/2,1/2]$ (the latter shown by vertical dotted lines). Spin densities $\langle \sigma_{1,2,3}\rangle$ 
	computed for the eigenfunctions  for the lowest (b) and the second-lowest (c) energy bands with $M=N=1$ at $k=0$. 
	Here   $\Lambda=3$, $\Omega_0=2$, $\alpha=2$, and $\theta = \pi/3$.}
	\label{fig:periodic}
\end{figure}

Before we proceed with our main results, it is instructive to compare the band structure corresponding to two least accurate BRAs in (\ref{eq:RA}).  For    $M/N=1$  the periods of both lattices are equal to $\pi$.   In Fig.~\ref{fig:periodic}(a) we compare   few lowest energy bands of the resulting  $\pi$-periodic Hamiltonian with the band structure of the  $2\pi$-periodic system obtained for $M/N=3/2$. We observe the standard splitting of the bands of the $\pi$-periodic lattice into the $N=2$ \emph{minibands}. In Fig.~\ref{fig:periodic}(b,c) we show the distribution of the spin densities $\langle \sigma_j\rangle=\bpsi^\dag\sigma_j\bpsi$ ($j=1,2,3$) at $k=0$.   

Although the eigenvalue problem (\ref{eigen}) with the periodic boundary conditions (\ref{pbc}) has purely discrete spectrum $E_\nu^{(N)}$ ($\nu=1,2,\cdots$), defining the  difference between the adjacent energies, $\Delta_\nu^{(N)} = E_{\nu+1}^{(N)}-E_\nu^{(N)}$ and bearing in mind the periodic continuation described above, one can employ the terminology of periodic potentials, considering `gaps' (relatively large $\Delta_\nu^{(N)}$) and `minigaps' (relatively small $\Delta_\nu^{(N)}$).

\begin{figure}
	\begin{center}
			\includegraphics[width=0.999\columnwidth]{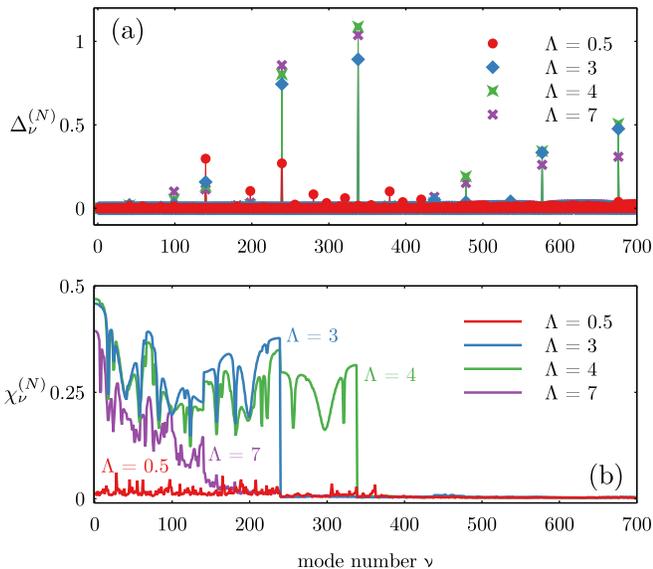}%
	\end{center}
	\caption{(a) Difference $\Delta_\nu^{(N)}$ between the adjacent  eigenenergies \textit{vs.} the mode number $\nu$ for several values of the constant Zeeman splitting $\Lambda$. (b) IPR $\chi_\nu^{(N)}$ for the same eigenvalues. In this figure $N=169$ ($M/N=239/169$), $\theta=\pi/3$, $\Omega_0=2$, and $\alpha=2$.  For each value of $\Lambda$, the corresponding plots    are obtained from   $700$ smallest eigenvalues. }
	\label{fig:gaps}
\end{figure}

Let us choose one of the BRAs from (\ref{eq:RA}), say, $M/N = 239/169$. In view of the variety of the parameters, we first focus on the effect of the  increasing constant Zeeman splitting $\Lambda$ with all other parameters being fixed. For few different values of $\Lambda$  we have computed several hundreds of smallest eigenvalues $E_\nu^{(N)}$. The eigenvalue problem (\ref{eigen})--(\ref{pbc}) has been   solved numerically using the Floquet-Fourier-Hill-type method \cite{Deconinck,Yang}.
In Fig.~\ref{fig:gaps}(a) we plot the differences $\Delta_\nu^{(N)}$ for several values of constant Zeeman field $\Lambda$. A distinctive pattern that can be observed from this plot indicates that, irrespectively of the value of $\Lambda$, the  gaps are situated between the modes with certain numbers determined by the BRAs (they are discussed below in Sec.~\ref{sec:memory}).  

\subsection{Localized modes and mobility edge}

 
Next we examine the localization, the existence of a ME, and spatial distribution of localized modes. In Fig.~\ref{fig:gaps}(b) we plot the IPR $\chi_\nu^{(N)}$ {\it versus} the mode number for several values of the constant Zeeman field $\Lambda$. At $\Lambda=0$, all eigenfunctions are delocalized (with IPR $\chi_\nu^{(N)}$ not exceeding 0.02, not shown in Fig.~\ref{fig:gaps}), {which can be partially explained by an additional symmetry that the system acquires at $\Lambda=0$ (see below in this subsection).}
For sufficiently small values of $\Lambda$ [see $\Lambda=0.5$ in Fig.~\ref{fig:gaps}(b)], the IPR remains small for all eigenfunctions. At larger values of $\Lambda$ a sharp ME emerges [see $\Lambda=3$ and $4$ in Fig.~\ref{fig:gaps}(b)]  which separates a fraction of localized modes with lower energies from the rest of the spectrum.  Comparing the two panels in Fig.~\ref{fig:gaps}, we observe that the location of the ME (when  the latter exists) coincides with the position of one of the gaps in the discrete spectrum (corroborating with previous studies on the one-component Schr\"odinger equation with quasiperiodic potential~\cite{Palencia}).  This implies that the chosen BRA determines not only the position of the gaps in the spectrum of eigenenergies, but also the number of localized eigenstates. In  Fig.~\ref{fig:gaps} this number is equal to 239 and 338 for $\Lambda=3$ and $4$, respectively. In  the meantime, the further increase of $\Lambda$,  {formally to the limit $\Lambda\to\infty$,} results in the degradation of  sharp ME and in the general decrease in the values of the IPR [see $\Lambda=7$ in Fig.~\ref{fig:gaps}(b)]. This is a manifestation of the Paschen-Back, alias nonlinear Zeeman, effect~\cite{PaschenBack,LL,BJ} (we notice that delocalization of particles caused by strong random SOC was recently described in~\cite{MSK}). {Large  Zeeman field $\Lambda$ results in a strong imbalance between the components (for large positive $\Lambda$ one has $|\psi_2|\gg |\psi_1|$), which means that the  effect of the SOC lattice becomes essentially perturbative, and the behavior of the system is dominated by the periodic Zeeman lattice which  alone is not sufficient   for the localization.}

We notice that the dimensionless $x$ is measured in the units of $\lambda/\pi$ where $\lambda$ is the  dimensional physical period of the Zeeman lattice, while the lattice amplitude is measured in the units $2E_r$ where $E_r=\hbar^2\pi^2/(2\lambda^2 m)$ is the recoil energy ($m$ is the atomic mass). Thus $\Omega_0=2$ corresponds to a lattice having amplitude $2E_r$, i.e., not too deep). The length $N \pi$ corresponds to $500\,\mu$m for $N=169$ and for the physical period of $\lambda=3\,\mu$m.

The fact that all eigenstates { shown in Fig.~\ref{fig:gaps}}  are delocalized for  zero and small values of the constant Zeeman field $\Lambda=0$ can be, to some extent, explained by analyzing symmetries of the system. We notice that for any $\Lambda$ both the quasiperiodic Hamiltonian $H$ and its superlattice  approximation $H^{(N)}$ feature {a time-reversal symmetry} $\sigma_3\cK$ ($\cK$ is the operator of complex conjugation): $[H,\sigma_3\cK]=[H^{(N)},\sigma_3\cK]=0$, meaning that any non-degenerate state is $\sigma_3\cK$-symmetric, and thus can be represented in the form $\bpsi=(\psi_1(x),i\psi_2(x))^T$ where $\psi_{1,2}(x)$ are real.  At the same time, for $\Lambda=0$ the Hamiltonian  $H^{(N)}$ with lattices  given by (\ref{eq:a(x)}) and (\ref{eq:Omega(x)}) acquires an additional symmetry. Introducing the translation over the half-period $T^{(N)}\colon x\to x+\pi N/2$, we obtain that  $[H^{(N)}, \sigma_1 T^{(N)}]=0$ for even $N$ and $[H^{(N)}, \sigma_1 T^{(N)}\cK]=0$ for odd $N$. This additional symmetry implies   that any localized state that exists at $\Lambda=0$  \rev{is  generically  degenerate and  consists of two peaks spatially separated by the half-period  $L^{(N)}/2$.} 


At the same time, the  half-period translation  symmetry does not completely forbid the localization at $\Lambda=0$: we found that the simultaneous increase of SOC lattice and Zeeman lattice depths (approximately starting with $\Omega_0\gtrsim 4$ and $\alpha \gtrsim 4$) eventually enables the existence of 
states composed of two localized peaks   spatially separated by the half-period \rev{(more precisely, there are two peaks in each component, $\psi_1$ and $\psi_2$, of the spinor eigenstate $\bpsi$)}. Localization for a  small but nonzero $\Lambda$ is also illustrated in Fig.~\ref{fig:Omega0}(c). 

\begin{figure}
	\begin{center}
		\includegraphics[width=0.99\columnwidth]{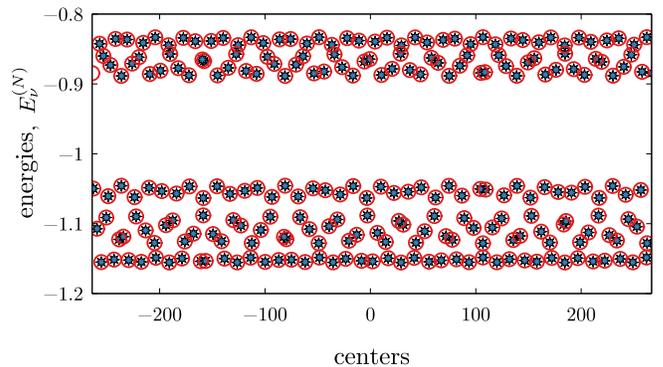}
	\end{center}
	\caption{Centers $\mathfrak{X}_\nu^{(N)}$  (red circles) and $\gamma_\nu^{(N)} N/2$ (blue stars) for all modes with energies below the ME.   In this figure $N=169$ ($M/N=239/169$), $\Lambda=3$, $\theta=\pi/3$, $\Omega_0=2$, and $\alpha=2$. There are 239 localized modes below the ME.}
	\label{fig:X}
\end{figure}

As established above in Eq.~(\ref{x-Zak}), the periodic continuation of the superlattice Hamiltonian over the entire real axis provides a  connection between the  spatial distribution of localized eigenstates and   the Zak phases of the bands of the corresponding superlattice. In order to illustrate this observation, in Fig.~\ref{fig:X} we compare the values $N\gamma_\nu^{(N)}/2$ computed numerically from the superlattice band spectrum with the centers  of localized states obtained directly by substituting the numerically found eigenvectors of (\ref{eigen}) into the   definition in  Eq.~(\ref{eq:Xp}).  The validity of  approximation (\ref{x-Zak}) is verified for  all localized modes, except for a few states   situated at the boundaries of the interval $I^{(N)}$. 

\subsection{Memory effect}
\label{sec:memory}

\begin{figure}
	\begin{center}
		\includegraphics[width=0.99\columnwidth]{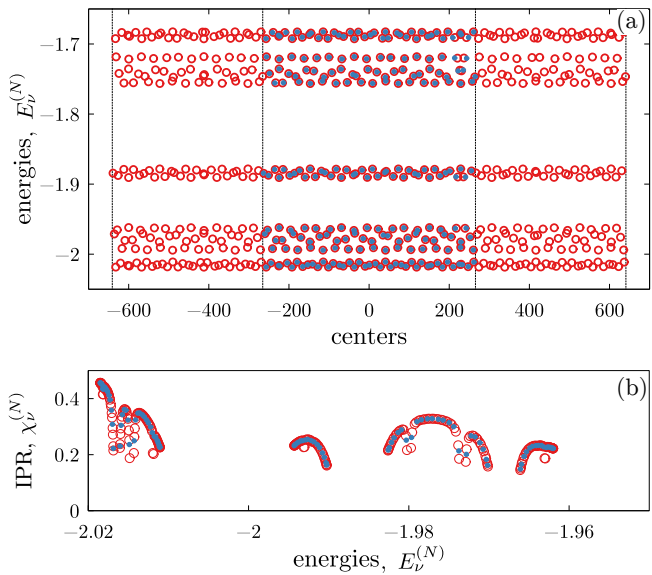}
	\end{center}
	\caption{(a) Centers $\mathfrak{X}_\nu^{(N)}$  for all modes with energies below the ME for two rational approximations: $M/N=239/169$ (filled blue  circles) and $M/N=577/408$ (open red circles). There are 239 and 577 localized modes below the ME, respectively. Vertical dotted lines correspond to the edges of the intervals $I^{(169)}$ and $I^{(408)}$,  $I^{(169)} \subset I^{(408)}$. \rev{(b) Dependence of IPR $\chi_\nu^{(N)}$ on energies $E_\nu^{(N)}$ for the two rational  approximations. Only the low-energy eigenstates $E_\nu^{(N)}<-1.95$ are shown in this panel.} In this figure, $\Lambda=5$,  $\Omega_0=2$, $\theta=\pi \sqrt{3}$,  and $\alpha=2$.  }
	\label{fig:nested}
\end{figure}

Figure~\ref{fig:X}  reveals several other interesting traits. For the chosen parameters, the localized modes are clustered in the two lowest bands, and the centers of modes form {a quasiperiodic} pattern in each band. The energy distribution of the modes is not uniform within each band: the localized modes tend to accumulate near the energy band edges revealing expectable increase of the density of states near the band edges. At the same time, the distribution of the centers of the localized modes along the interval $I^{(N)}$ is nearly uniform. This suggests an intuitive explanation to the fact that the spectrum computed for the given BRA `remembers' some information about the previous (i.e.,  less accurate) BRAs.   Indeed, let $M_1/N_1$ and $M_2/N_2$ be two BRAs with   $N_1<N_2$, i.e., the second BRA is more accurate. Then  the corresponding superlattice period $I^{(N_2)}$ can be represented as  $I^{(N_2)}=I_-\cup I^{(N_1)}\cup I_+$, where $I_-$ and $I_+$ are intervals $[-\pi N_2/2,-\pi N_1/2)$ and $[\pi N_1/2, \pi N_2/2)$, respectively. Both intervals $I^{(N_1)}$ and $I^{(N_2)}$ are uniformly covered by localized states. At the same time, if $N_1$ and $N_2$ are large, then the states inside the interval $I^{(N_1)}$ located sufficiently far from its boundaries, are weakly affected by the replacement of $\Omega^{(N_1)}(x)$ by $\Omega^{(N_2)}(x)$, because  $\delta\Omega(x)=\Omega^{(N_2)}(x)-\Omega^{(N_1)}(x)
\sim  \left(\frac{M_2}{N_2}-\frac{M_1}{N_1}\right)\Omega_0x$ is small enough. In other words, passing from the less accurate BRA  to the more accurate one,   one does not affect significantly the states localized within  $I^{(N_1)}$, i.e., obtained under the less accurate approximation. Notice that $|\delta\Omega(x)|$ is a very small quantity: after a few first approximations, say for $N_1=70$ and $N_2=169$, even at the boundaries of $I^{(N_1)}$ one has $\delta\Omega(x)\approx 9\cdot 10^{-3}\Omega_0$. Thus, the localized states in the interval $I^{(N_2)}$ can be viewed as the weakly deformed states of the previous   approximation in $I^{(N_1)}$ complemented by the `new' states which are located mainly in the intervals $I_\pm$. Since the localized states are uniformly distributed,   the number of localized states ($n^{(N_{1,2})}$) for BRAs with $N_1$ and $N_2$ are interrelated as
\begin{equation}
n^{(N_2)} \approx  n^{(N_1)} +  n^{(N_1)}\frac{N_2-N_1}{N_1} =  n^{(N_1)} \frac{N_2}{N_1}.
\end{equation}
In other words, considering a sequence of BRAs $N_1<N_2<N_3<...$ one obtains a nested structure of $I^{(N)}$-intervals: $I^{(N_1)}\subset I^{(N_2)}\subset I^{(N_3)}\subset \cdots$. The `memory' that $N_j$th approximation has about its $N_{j-1}$th predecessor consists of the modes  located inside $I^{(N_j)}$ and hence inside $I^{(N_{j-1})}$, as well, i.e., of the modes that belong to both intervals.   To illustrate this `memory effect', in Fig.~\ref{fig:nested} we juxtapose the  energies and centers of localized modes obtained under two subsequent BRAs with $N_1 = 169$ an $N_2=408$.   In Fig.~\ref{fig:nested} we observe that  within the smaller interval $I^{(169)}$ the centers and energies  computed for both BRAs coincide for almost all  localized modes, except for  few modes situated near the boundaries of this interval,  i.e., the more accurate BRA retains the information about the localized modes that exist under the previous BRA.

The memory effect can also be observed from the  position of the gaps in the miniband spectra plotted in Fig.~\ref{fig:gaps}(a) (for the stationary Schr\"odinger equation with an incommensurate bichromatic lattice potential this phenomenon was discussed in~\cite{Diener}). Trying to understand why the largest gaps $\Delta_\nu^{(N)}$ occur   exactly at certain positions $\nu$,  let us go back to the band structure of the $\pi$-periodic lattice [shown in Fig.~\ref{fig:periodic}(a)] that corresponds to the least accurate BRA with $N=M=1$. Passing to a more accurate BRA with $M,N>1$, each band of the former lattice splits into $N$ minibands (this is illustrated in Fig.~\ref{fig:periodic}(a) for $M/N=3/2$).  It is natural to expect that the   emerging minibands are  situated close to bands of the original lattice. This anticipation would imply that  the largest gaps between the minibands, i.e., the largest values of   $\Delta_\nu^{(N)}$ correspond to   $\nu=N,\,2N,\,3N, \ldots$.  However, already from Fig.~\ref{fig:periodic}(a) it is evident that this first expectation is only partially correct:  some of the minibands are   situated close to each other even though they emerge from different bands of the original lattice. Nevertheless, the expected pattern still manifests  itself: analysing the location of the gaps in Fig.~\ref{fig:gaps}(a),  one can see that the largest  value of $\Delta_\nu^{(N)}$ appears at   $\nu=338$ thus corresponding to $169+169$ (i.e., to the chosen $2N$). In the meantime, the second largest gap in Fig.~\ref{fig:gaps} appears at $\nu=239$, i.e., the relation  $239=169+70$ is verified. Other large  gaps are situated at   $\nu=99,\, 140, \, 478,\,  577, \, 676$.
Each of these numbers can be represented as a sum of two (or four) denominators in the sequence (\ref{eq:RA}):  $99=70+29$, $478=2(70+169)$, $577=3\cdot 169+ 70$, $676=4\cdot 169$. 
Thus, in this picture the positions of the gaps in the discrete spectrum are determined not only by the particular BRA, that is  $M/N = 239/169$ chosen in the numerical simulations,  but also by the  previous   BRAs in the series (\ref{eq:RA}). In other words, the spectrum obtained for some particular BRA preserves certain information about the previous, less accurate approximations in the sequence (\ref{eq:RA}).

\subsection{Global picture}

\begin{figure}
	\begin{center}
			\includegraphics[width=0.999\columnwidth]{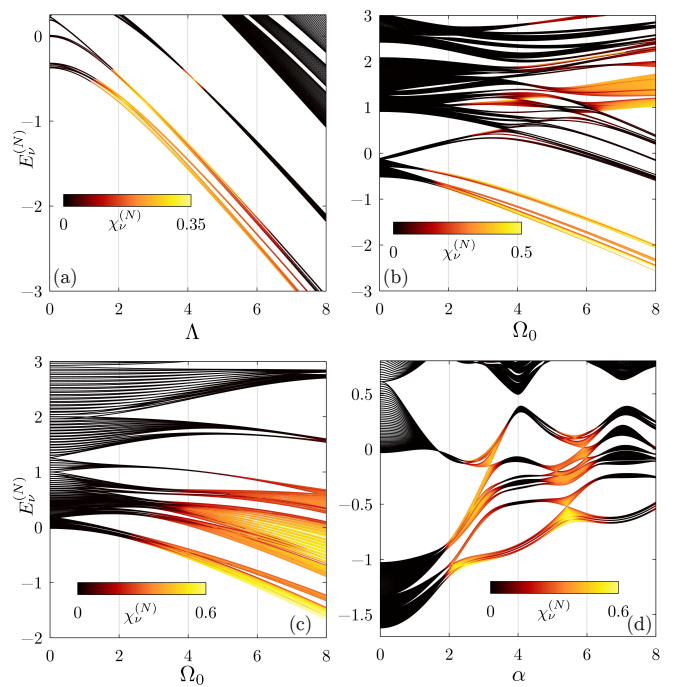}%
	\end{center}
	\caption{Pseudocolor plot of IPR {\it vs.} the amplitude of the constant Zeeman splitting $\Lambda$  (a),   the amplitude of the Zeeman lattice $\Omega_0$ (b, {c}), {and the amplitude of the SOC lattice $\alpha$ (d).} Other parameters are chosen as $\alpha=2$, $\Omega_0=2$ in (a), $\alpha=2$, $\Lambda=2$ in (b), {$\alpha=4$, $\Lambda=0.1$ in (c), and $\Lambda=3$, $\Omega_0=2$ in (d)}. All panels are  obtained for  $\theta=\pi/3$, and  $M/N=239/169$ (a,b,d) and $M/N=99/70$ (c).}
	\label{fig:Omega0}
\end{figure}

The above study was focused on isolated values of the constant Zeeman splitting $\Lambda$. Now we  fulfill a more thorough examination of    the eigenspectrum of problem (\ref{eigen}) scanning a finite interval of $\Lambda$.
In Fig.~\ref{fig:Omega0}(a) we plot the computed points $(\Lambda, E_\nu^{(N)})$. The pseudocolor represents the value of IPR for each computed eigenvector. The obtained general picture agrees with the previous considerations. Namely, we observe that for small and large values of $\Lambda$ the spectrum is poorly localized {i.e., characterized by relative small IPRs corresponding to the wavepacket widths comparable with the size of the system $L^{(N)}$ }. 
A sharp ME, i.e., the one separating energies of deloclaized states and the states whose localization domain is much less than the size of the system exists for $2 \lessapprox   \Lambda \lessapprox 6$). The location of the ME coincides  with the upper edge of the higher band in which the states are localized.

In Fig.~\ref{fig:Omega0}(b) we show a preudocolor diagram obtained for a situation when the value of the constant Zeeman field $\Lambda$ is fixed, but the amplitude of the Zeeman lattice $\Omega_0$ increases gradually departing from zero.  As one can expect, all states are delocalized in a Zeeman lattice of zero and  small amplitudes ($\Omega_0\lesssim 2$). The localization of states from the lowest gaps gradually enhances with the increase of $\Omega_0$. Moreover, for sufficiently large $\Omega_0$, one observes several MEs emerging between the groups of energies associated with localized and delocalized eigenstates. In this last case clusters of localized states have energies bigger than those of delocalzied states [in Fig.~\ref{fig:Omega0}(b) this is clearly seen for $\Omega_0\gtrsim 6$]. 

In Fig.~\ref{fig:Omega0}(c)  we illustrate the possibility of localization in the Zeeman field with small mean value $\Lambda=0.1$. We observe that in this case the localization is also possible, but requires the presence of deeper  lattices  as compared to the localization under intermediate values of $\Lambda$ [compare the values of $\alpha$ and $\Omega_0$ in Fig.~\ref{fig:Omega0}(a) and Fig.~\ref{fig:Omega0}(c)]. Finally, in Fig.~\ref{fig:Omega0}(d) we present   localization diagram for the increasing SOC lattice amplitude $\alpha$. Localization is observed only for a finite interval of the SOC strength and is characterized by inhomogeneous dependence of the ME on $\alpha$: for the parameters of Fig.~\ref{fig:Omega0} (d) the localization domain is $2\lesssim\alpha\lesssim 6.5$ while the ME is much lower for the central part of this interval. 


\subsection{Degenerate modes}
\label{sec:double}

In the generic situation the   spectrum of (\ref{eigen}) is non-degenerate, i.e., $E_\nu^{(N)}<E_{\nu+1}^{(N)}$ for all (or almost all) $\nu$.   However, for special values of the phase shift $\theta$ [see Eq.~(\ref{eq:Omega(x)})] between the SOC and Zeeman lattices, 
certain BRAs may enable additional symmetries that result  in  degeneracies, i.e., in large number of double eigenenergies $E_\nu^{(N)}=E_{\nu+1}^{(N)}$ emerging in the spectrum. To look for  an additional symmetry, let us consider the phase shift of the form $\theta = Q_1 \pi/ Q_2$ where $Q_{1,2}$ are coprime integers.  Then  one can observe that the transformation  $x\to P\pi/2 - x$, where integer $P$ satisfies the following condition 
\begin{equation}
	\label{eq:symmetry}
	2\frac{Q_1}{Q_2} + P\frac{M}{N}  = \mathrm{odd~integer},
\end{equation}
transforms the superlattice Hamiltonian $H^{(N)}$ with the SOC and Zeeman lattices given by (\ref{eq:a(x)}) and (\ref{eq:Omega(x)}) either to itself (if $P$ is odd) or to its complex conjugate (if $P$ is even). If  $\bpsi_{\nu}^{(N)}(x)$ is eigenvector corresponding to the energy $E_\nu^{(N)}$, then, depending on the parity of  $P$,   $\bpsi_{\nu}^{(N)}(P\pi/2 - x)$ or $\cK{\bpsi}_{\nu}^{(N)}(P\pi/2 - x)$   
is also an eigenstate corresponding to the same eigenenergy  $E_\nu^{(N)}$. If the original and transformed eigenvectors are linearly independent (which is true for   most of the modes), then $E_\nu=E_{\nu+1}$ is a double eigenvalue which has a two-dimensional invariant subspace  spanned by the found eigenvectors.

Considering  (\ref{eq:symmetry})   as an equation for an unknown $P$ with $Q_1/Q_2$ and $M/N$ being fixed, we observe that this equation not   always has a solution.   For example, for $Q_1/Q_2 = 1/3$ (this is the case considered above) the BRA  $M/N=239/169$ does not allow for integer solution $P$. However, for the next BRA from the sequence (\ref{eq:RA}), i.e., for $M/N=577/408$, one finds a solution $P=136$. Thus, different  BRAs can be `nonequivalent' with respect to this spontaneous symmetry.  In the case when the symmetry is present, we have verified numerically that a large number of double eigenenergies  $E^{(N)}_\nu=E^{(N)}_{\nu+1}$  emerge in the spectrum. Each genenerate eigenvalue is associated with a   two-peaked eigenvector, and the distance between the two peaks is equal to $P\pi/2$.


\section{Dynamics}
\label{sec:dynamics}

Finally we briefly discuss the evolution of a time-dependent spinor $\bPsi(x,t)$ described by the Schr\"odinger equation $i\bPsi_t(x,t) = H^{(N)} \bPsi(x,t)$ \rev{considered subject to the periodic boundary conditions (\ref{pbc}). Aiming at a qualitative preliminary study} here we present several explicit simulations of the dynamics which illustrate the effect that the presence of localized modes has on  the temporal  dynamics. The time-dependent equation has been integrated with the $\sigma_3\cK$-symmetric initial condition  
$\bPsi(x,t=0) = (1, i)^T\Psi(x)$, where $\Psi(x)$
is a real-valued Gaussian wavepacket (the IPR of the initial pulse was  
$\chi(t=0)=0.2$). Since the localized modes are distributed uniformly on the interval  $I^{(N)}$, the qualitative dynamics does not depend significantly on the position of the initial wavepacket. We have created the initial state at the center of the  interval $I^{(N)}$.  The numerical solution $\bPsi(x,t)$ has been then used to find the time-dependent IPR $\chi(t)$ and the mean width.

\begin{figure}
	\begin{center}
		\includegraphics[width=0.999\columnwidth]{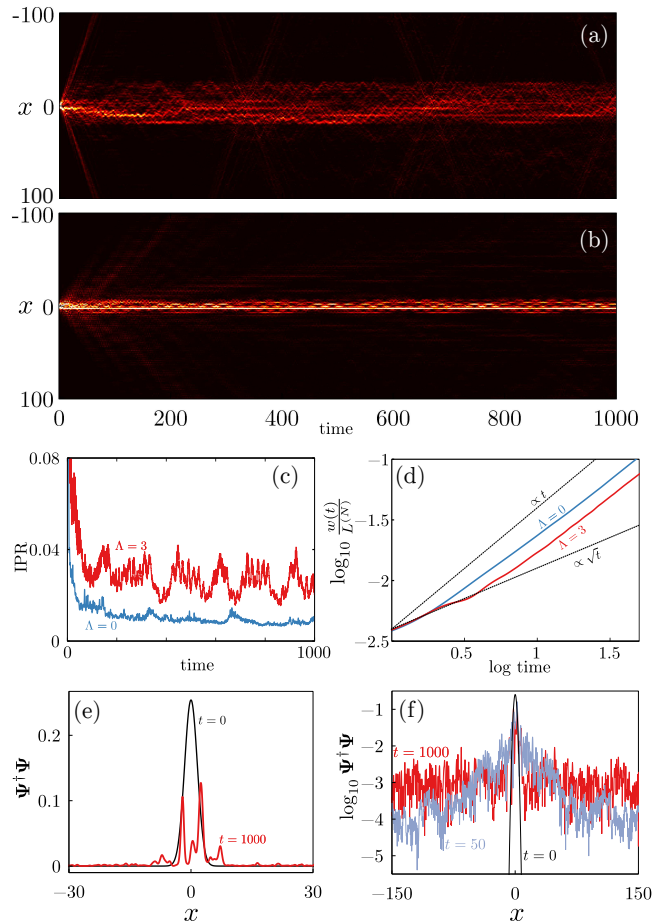}
	\end{center}
	\caption{ (a,b) Pseudocolor plots of the temporal evolution $\bPsi^\dagger\bPsi$ below and above the delocalization-localization transition ($\Lambda=0$ and $\Lambda=3$ in the upper and lower panels, respectively).  Only the central part $x\in[-100,100]$ of the entire interval $I^{(N)}$ is shown in  the plots. (c) The corresponding evolution of the IPR $\chi(t)$. (d) $\log$-$\log$ plot of the widths { in units of the full superlattice period $L^{(N)}=\pi N$}, \rev{ plotted for $t\in (0,50)$}. Black dotted lines correspond to the square-root ($\propto \sqrt{t}$) and  linear ($\propto {t}$) laws. {Panels (e,f) show $\bPsi^\dag \bPsi$ at $\Lambda=3$  and different moments of time $t$  in linear and log scales.} In this figure     $M/N=239/169$, $\Omega_0=2$, $\alpha=2$,   $\theta=\pi/3$. }
	\label{fig:dyn}
\end{figure}

In the presence of localized states, the initial wavepacket excites those situated along  the width of the distribution $\Psi(x)$, leading to a nondispersing part of the wavepacket. Additionally, higher-energy extended sates are excited,  which results in partial dispersion. \rev{The qualitatively different dynamics in the extended and localized phases are } illustrated in Fig.~\ref{fig:dyn}(a,b).  Comparing the pseudocolor plots of $\bPsi^\dagger\bPsi$ we observe that in the case when the localized modes are present in the spectrum, the atom is dominantly localized in the central region [Fig.~\ref{fig:dyn}(b)]. Meantime, even in the absence of localized states, the dispersion of the wavepacket is slowed down by the quasiperiodicity [Fig.~\ref{fig:dyn}(a)].  For characterization of the evolution of the localized modes the time dependent IPR  appears to be the most appropriate quantity [exemplified in Fig.~\ref{fig:dyn}(c)], while dispersive spreading at early stages of dynamics is more adequately described by  the mean-square width of the wavepacket $w(t)  =\left(\int_{I^{(N)}}[x-x_0(t)]^2\bPsi^\dag \bPsi dx\right)^{1/2}$, where $x_0(t) = \int_{I^{(N)}}x \bPsi^\dag  \bPsi dx$ shown in Fig.~\ref{fig:dyn}(d). In the presence of localized modes, time evolution of IPR reveals relatively strong oscillations [see the red line in Fig.~\ref{fig:dyn}(c) which shows a superposition of fast and slow oscillations of $\chi(t)$]. Taking into account that the distribution of energies of localized states is nonuniform, the oscillations can be viewed as manifestation of quantum revivals, typical for a trapped quantum particle.  

Figure~\ref{fig:dyn}(d) shows the  log-log plot for \rev{short-time} evolution of the mean width $w(t)$ of the solutions. \rev{We observe the  $w(t)\propto t^\beta$ law with $\beta$ varying between $\beta=1/2$ at the beginning of the evolution and $\beta=1$ at larger   times [see the dashed lines in Fig.~\ref{fig:dyn} (d)]. It is to be mentioned that the exponent $\beta$ in the nontrivial $t^\beta$ dependence is likely to be related to the fractal spectrum of the quasiperiodic Hamiltonian (\ref{hamilt}), as suggested by earlier studies~\cite{dispersion} (the investigation of such a relation, however, is left for the further investigation). } 

\rev{The atomic density distribution at $t=1000$ is illustrated in Fig.~\ref{fig:dyn}(e). Although the initial shape of the cloud at $t=1000$ is strongly modified, the localization domain remains nearly the same as at was at $t=0$. This is also confirmed in the logarithmic plot in Fig.~\ref{fig:dyn}(f), where we also show the intermediate distribution at $t=50$, when the spreading wavepacket still does not attain the boundaries of the spatial domain $L^{(N)}$ used in numerics. One verifies that while the boundaries do affect the decay of the wave-packet tails, they do not have significant impact on the atomic cloud distribution in the localization region located in center of the interval $L^{(N)}$, even at sufficiently large times.}

\section{Conclusion}
\label{sec:concl}

In this paper, we have considered properties of the atomic spinor under the effect of incommensurate    periodic spin-orbit-coupling lattice and  Zeeman lattice.
{When the constant Zeeman splitting is zero, the system acquires an additional symmetry which constraints the shape of  eigenfunctions and therefore inhibits the localization, i.e., requires stronger lattice depths for the localization to occur.} 
At sufficiently large constant Zeeman field all modes become delocalized  thus manifesting the well-known Paschen-Back effect. Using best rational approximations, the consideration of quasiperiodic Hamiltonian has been reduced to   an effective periodic superlattice. 
When the mobility edge exists,   the  exact number of localized eigenstates is determined  by the chosen rational approximation. Moreover, considering best rational approximations of different accuracy, we found that the more accurate approximation preserves certain information about the rougher one,   which we interpreted as a  `memory' effect. Furthermore, spatial positions of localized eigenmodes can be obtained from the Zak phases of minibands of the effective superlattice,  revealing the relation between the distribution of the modes in the space and the topology of the effective superlattice. In the presence of localized modes, simulations of evolution of initially localized wavepackets reveal the signature of quantum revivals.

The obtained results are directly applicable to noninteracting spin-orbit-coupled  Bose-Einstein condensates. It may be therefore relevant to extend the present study to the condensates with nonzero interatomic interactions, which is expected to    further enrich the host of features resulting from the combination of the two incommensurate lattices.    

\rev{We finally mention that, while the presented results of temporal simulations uncover a qualitatively different wavepacket dynamics   in the extended and localized phases, a comprehensive and accurate description of different stages of long-time  dynamics requires a separate thorough study.} 
 
\begin{acknowledgments}

\rev{We  are grateful to the    anonymous Referee for drawing our attention to Ref.~\cite{dispersion}.}
 The work of D.A.Z. was supported by the Foundation for the Advancement of Theoretical Physics and Mathematics ``BASIS'' (Grant No. 19-1-3-41-1) and Priority 2030 Federal Academic Leadership Program. 
V.V.K. acknowledges financial support from the Portuguese Foundation for Science and Technology (FCT) under Contracts PTDC/FIS-OUT/3882/2020 and UIDB/00618/2020.
\end{acknowledgments}

\end{document}